\documentclass{PoS}

\title{Exploring New Physics in the $C_7$-$C_{7'}$ plane}

\ShortTitle{Exploring New Physics in the $C_7$-$C_{7'}$ plane}

\author{\speaker{S\'ebastien Descotes-Genon}\\
Lab. de Physique Th\'eorique, CNRS/Univ. Paris-Sud 11 (UMR 8627),
91405 Orsay, France
}
\author{Diptimoy Ghosh\\
Tata Institute of Fundamental Research, Homi Bhabha Road, Mumbai 400005, India 
}
\author{Joaquim Matias and Marc Ramon\\
Universitat Autonoma de Barcelona, 08193 Bellaterra, Barcelona, Spain}

\abstract{We discuss the model dependence in the determination of the Wilson coefficient $C_7$ that governs the radiative electromagnetic decays of $B$ mesons in the Standard Model, by considering various extensions of the effective Hamiltonian used to describe such decays.
We include already measured observables like the branching ratios of $B \to X_s \mu^+ \mu^-$ and
$B \to X_s \gamma$, the isospin and CP asymmetries in $ B\to K^{*}\gamma $, as well as $A_{\mathrm{FB}}$ and $F_{\mathrm{L}}$ in $B \to K^* \ell^+\ell^-$, adding the LHCb measurements presented at this conference. 
We 
explore the constraints on $C_7,C_9,C_{10}$ as well as their chirality-flipped counterparts.
We also discuss the transverse asymmetry $A_{\mathrm{T}}^{(2)}$ which, once measured, may help to disentangle some of the scenarios considered.}

\FullConference{XXIst International Europhysics Conference on High Energy Physics\\
                 21-27 July 2011\\
                 Grenoble, Rh\^one-Alpes, France}

\begin{document}

\begin{figure}
\begin{center}
\includegraphics[width=7cm]{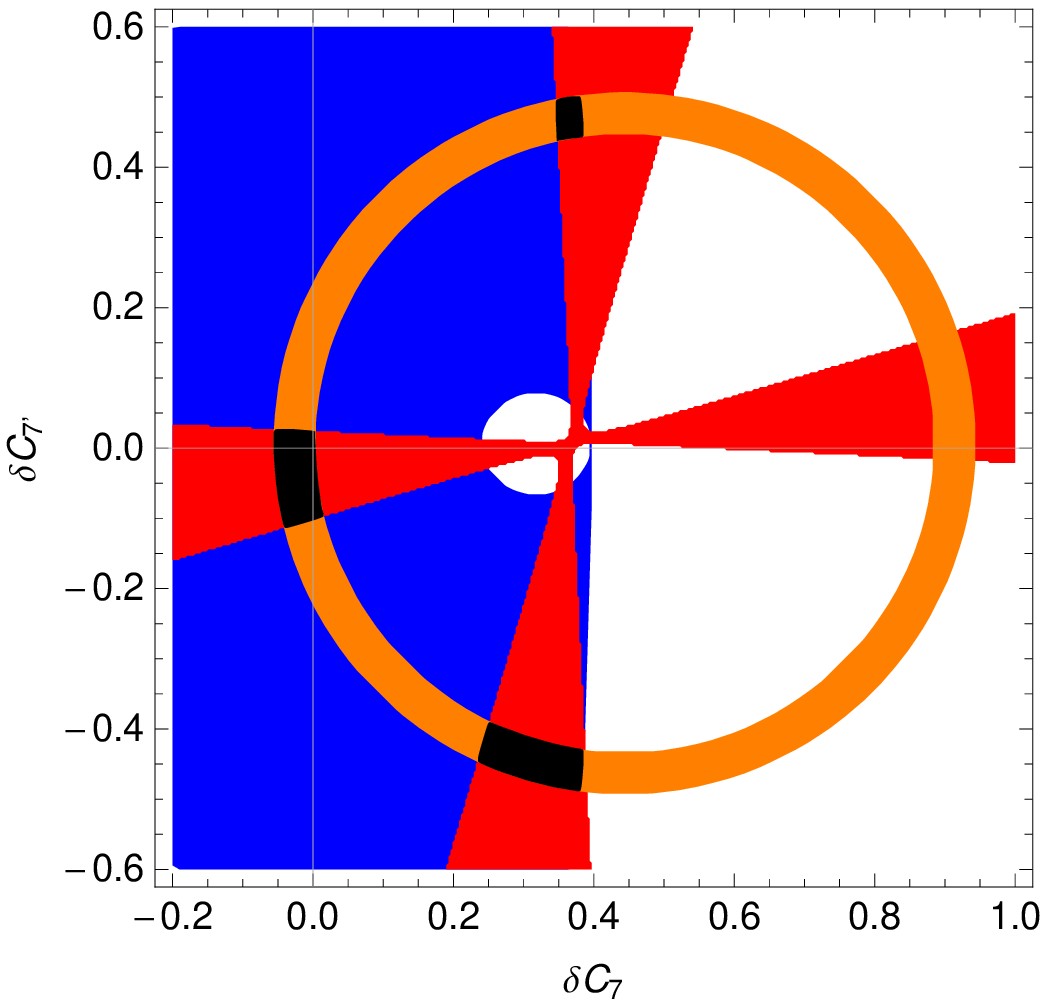}
\includegraphics[width=8cm]{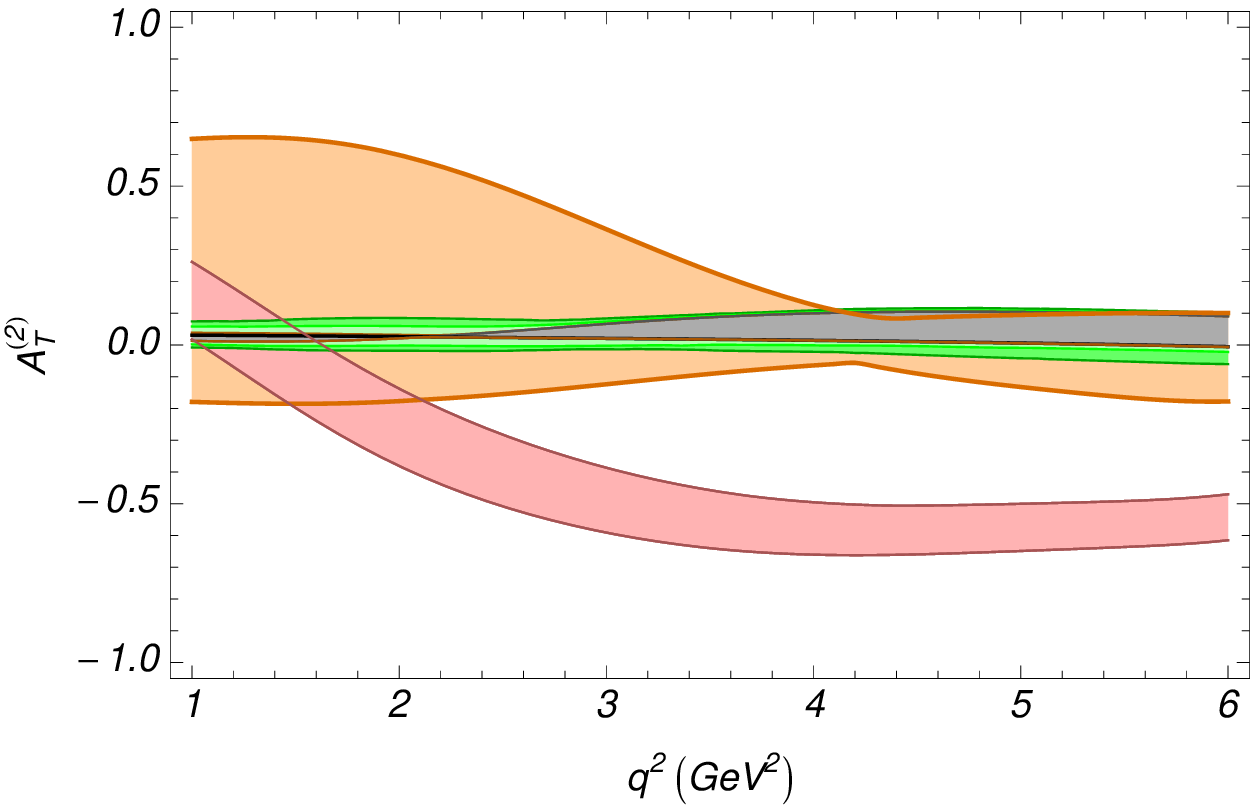}
\end{center}
\caption{On the left: class-I observables at $1\, \sigma$, $A_I$ (solid blue region with a white disk), $Br(B \to X_s \gamma$) (orange ring) and $S_{K^* \gamma}$ (red cross). The three  regions allowed by the intersection of the three constraints are depicted in black. The Wilson coefficients are taken at $\mu_b=4.8$ GeV, and $\delta C_i$ is the deviation with respect to the SM value [corresponding to $(\delta C_7,\delta C_{7^{\prime}})=(0,0)$].
 On the right: envelopes of the $q^2$-dependence for $A_{T}^{(2)}$ under Scenario A for the two regions allowed by class-I and class-III observables. The yellow (respectively pink) envelope corresponds to $(\delta C_7,\delta C_{7^{\prime}})$ in the magenta area close to the SM region (resp. in the circle in the lower ``inverted'' region) on the right (resp. left) panel of fig.~2.}\label{fig:C7C7p}
\end{figure}

The very good agreement between the Standard Model (SM) expectations and experimental data in flavour physics sets particularly stringent constraints on any model of New Physics (NP). A promising field to identify patterns of NP from experiment is provided by radiative (and dileptonic) $b\to s$ decays, as these loop processes, to be measured extensively at LHCb, have a potential sensitivity to phenomena beyond the Standard Model. 
In ref.~\cite{DescotesGenon:2011yn}, we proposed to focus on the electromagnetic operator ${\cal O}_7$ and its chirally-flipped counterpart ${\cal O}_{7^\prime}$, defined in the effective Hamiltonian approach, as tools to search for New Physics in a systematic way. These coefficients
play here a similar role to the $\bar\rho$ and $\bar\eta$ parameters in the studies of the unitarity triangle. $C_7$ and $C_{7^\prime}$ do not exhaust all the information that can be obtained concerning NP, exactly as $\bar\rho$ and $\bar\eta$ are not sufficient to describe the full structure of the CKM matrix, but they provide an interesting summary of the situation and a good starting point to investigate NP contributions.
Our {\it framework}  is defined by considering that NP enters in ${\cal O}_i$ with $i=7,9,10$ (electromagnetic and semileptonic operators), together with the chirally-flipped operators ${\cal O}_{i^\prime}$ with $i=7,9,10$. Within this framework, three \emph{scenarios A,B,C} correspond to switching on NP step by step: \emph{A)} NP affects the electromagnetic dipole operators ${\cal O}_7$, ${\cal O}_{7^\prime}$, \emph{B)} NP enters not only ${\cal O}_7$, ${\cal O}_{7^\prime}$, but also the semileptonic operators ${\cal O}_9$ and ${\cal O}_{10}$,
\emph{C)} all operators ${\cal O}_{7,9,10}$ and (chirally-flipped) ${\cal O}_{7^\prime,9^\prime,10^\prime}$ can receive NP contributions.  

We 
assume that NP enters only these operators, and that their Wilson coefficients are real. If no solution compatible with all constraints is found at the end of our analysis, within our defined framework, the next step consists in generalizing the framework to other operators. Accordingly, we classify our observables, chosen for their limited sensitivity to hadronic uncertainties and/or their important impact on the Wilson coefficients, in three {\it classes}:

\emph{1)} \emph{Class-I observables} mainly sensitive to ${\cal O}_7$ and ${\cal O}_{7^\prime}$, but not to ${\cal O}_{i=9,10,9^\prime,10^\prime}$:  the branching ratio of the inclusive radiative decay $B \to X_s \gamma$, as well as the isospin asymmetry ($A_I$) and the CP-asymmetry ($S_{K^* \gamma}$) of the exclusive decay $B\to K^*\gamma$. 

\emph{2)} \emph{Class-II observables} exclusively sensitive 
 to ${\cal O}_7$ and ${\cal O}_{7^\prime}$,  to semileptonic operators ${\cal O}_9$ and ${\cal O}_{10}$ and their chiral counterparts  ${\cal O}_{9^\prime}$, ${\cal O}_{{10}^\prime}$. Even within more general frameworks, only these operators occur in $A_{T}^{(2)}$,  an asymmetry defined from
an  uniangular distribution of $B\to K^*\ell^+\ell^-$~\cite{Kruger:2005ep}.

\emph{3)} \emph{Class-III observables} that are sensitive to all the previous operators, and in addition they may exhibit a sensitivity to NP contributions from other operators (scalars, tensors, chromomagnetic\ldots): this occurs for $Br(B \to X_s \ell^+ \ell^-)$ and observables from the angular distribution of $B \to K^*(\to K \pi) \ell^+ \ell^-$ (forward-backward asymmetry $A_{\mathrm{FB}}$, longitudinal polarisation $F_{\mathrm{L}}$).

For exclusive quantities, we work within QCD factorisation~\cite{QCDF} to simplify the analysis of the form factors, and we consider only averaged data over the low-$q^2$ region (invariant leptonic mass from 1 to 6 GeV$^2$).
In ref.~\cite{DescotesGenon:2011yn}, we provided semi-numerical formulae for our observables (see also this reference for the inputs and methods used), allowing us to exploit the experimental results available for these observables easily. In the present proceedings and in an upcoming addendum~\cite{DescotesGenon:2011yn}, we provide updated results including the LHCb results presented in this conference, shifting the inputs accordingly: $\tilde{A}_{FB}=0.33^{+0.22}_{-0.24}\to 0.04\pm 0.12$ and $\tilde{F}_L = 0.60^{+0.18}_{-0.19} \to 0.60 \pm 0.09$.
The reader may compare with ref.~\cite{DescotesGenon:2011yn} to identify the differences induced by these changes.

\begin{figure}
\includegraphics[width=4.99cm]{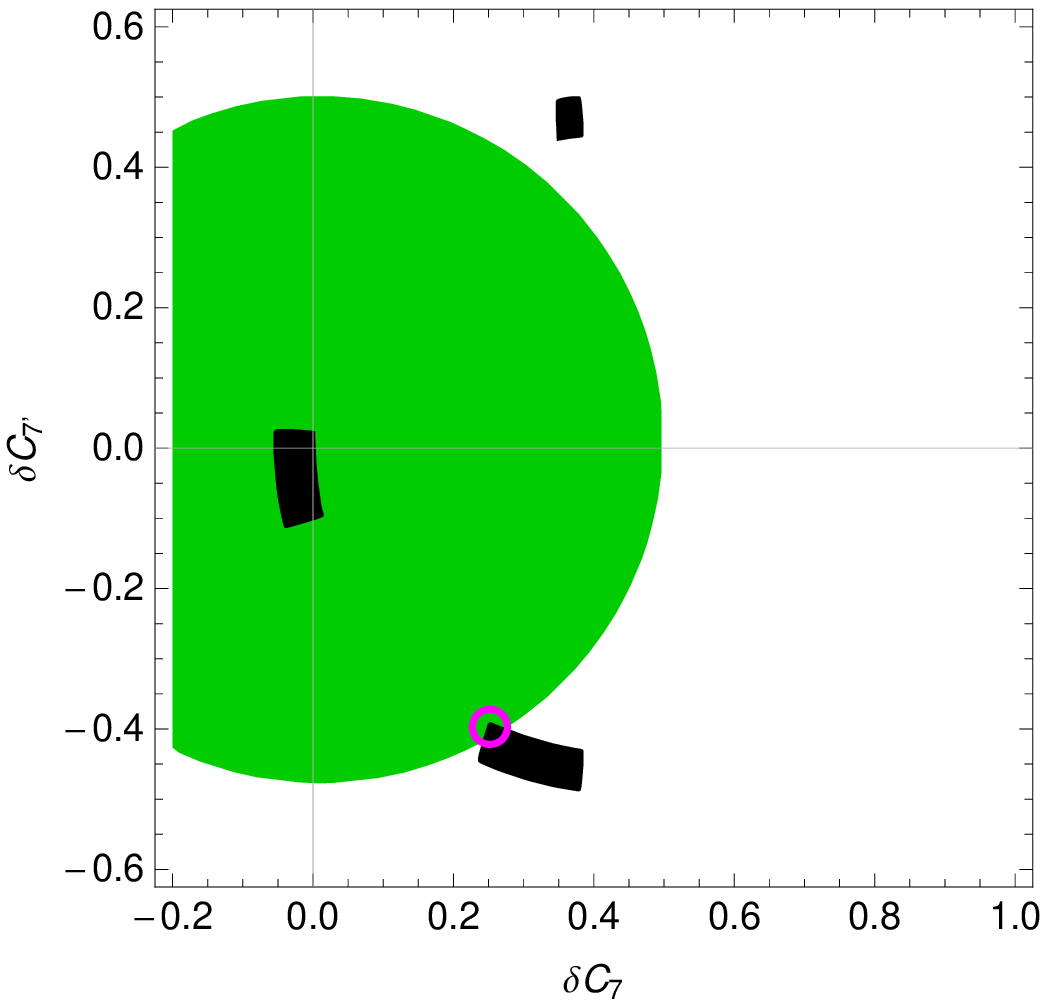}
\includegraphics[width=4.99cm]{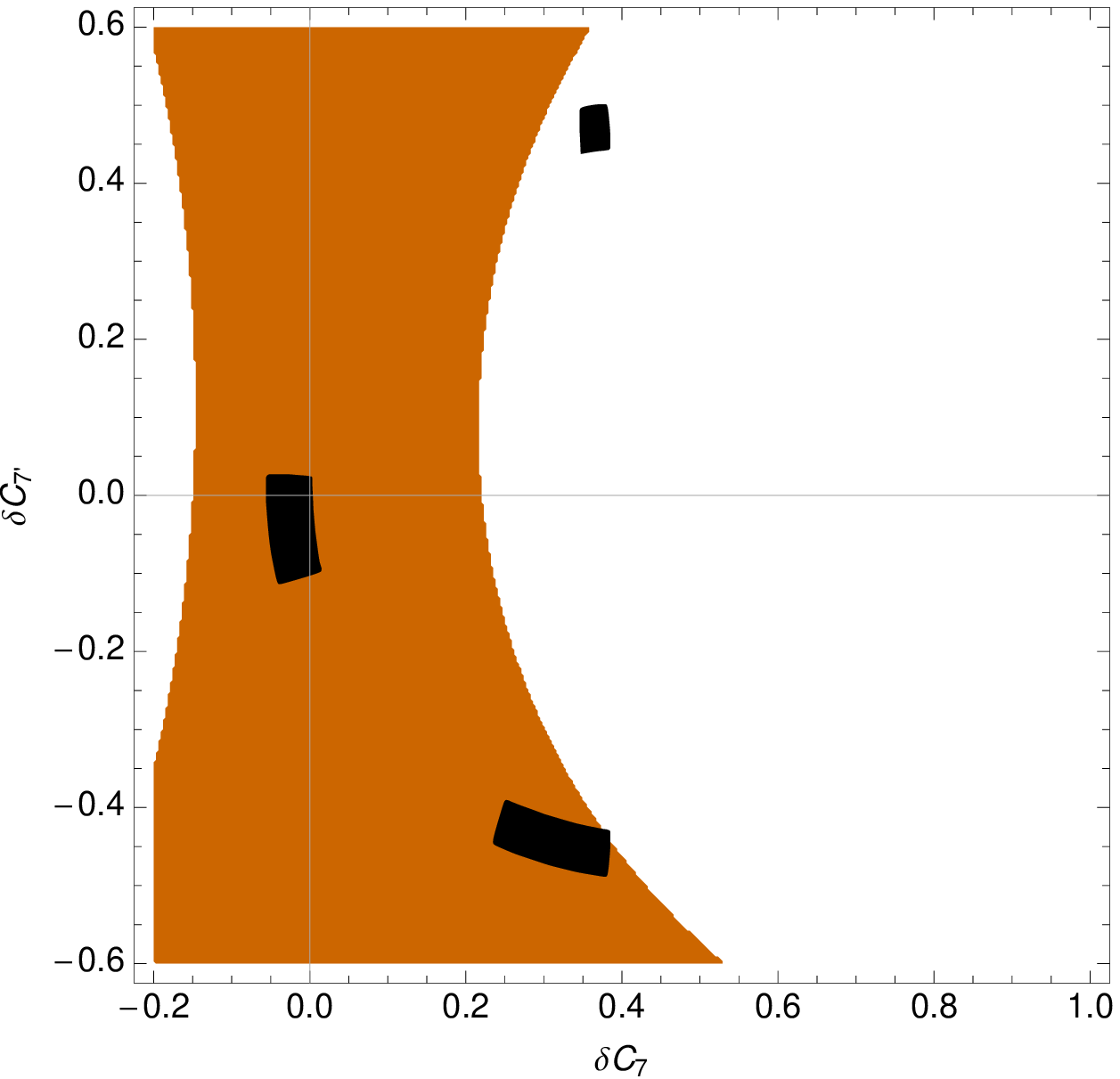}
\includegraphics[width=4.99cm]{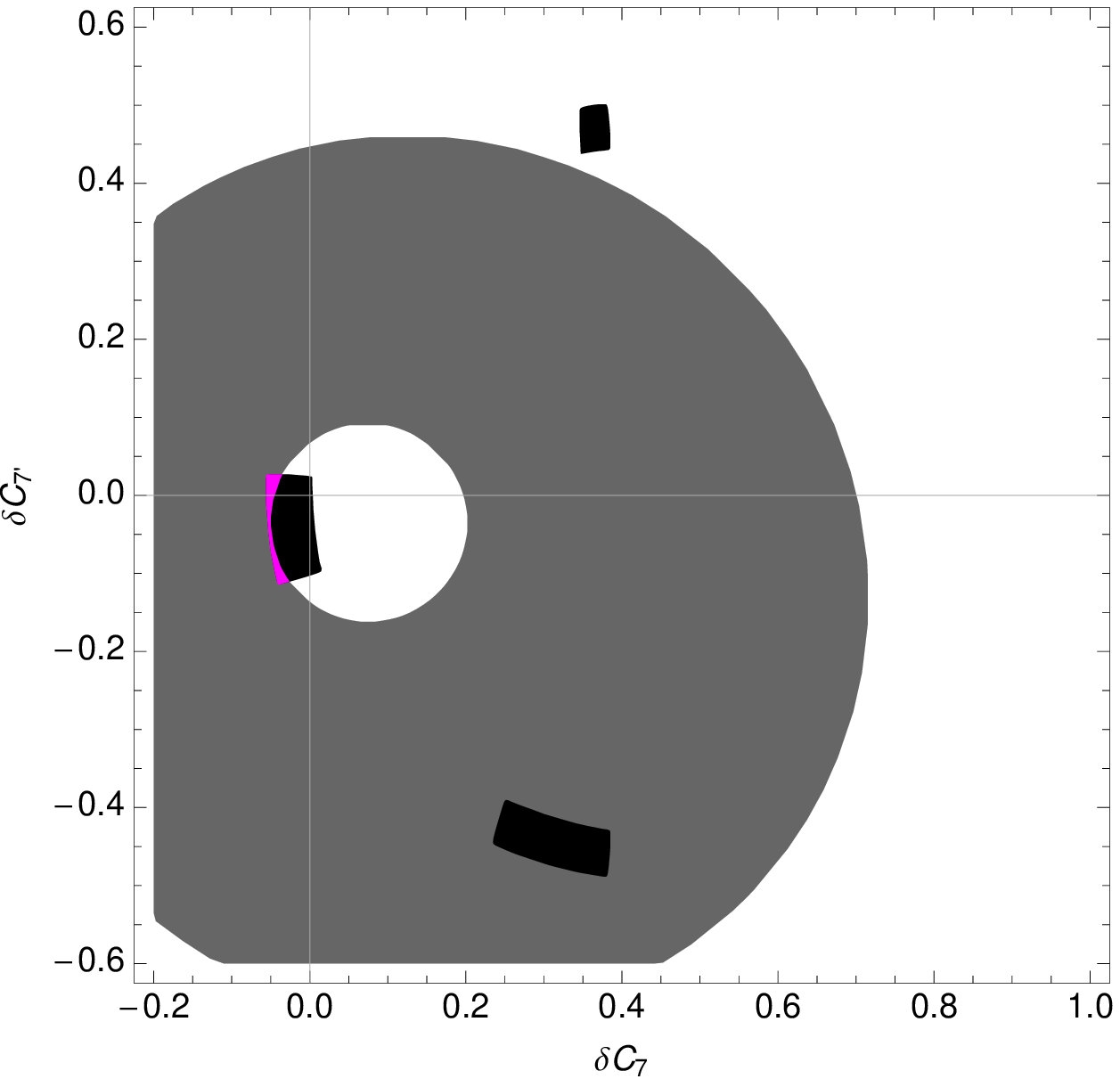}
\caption{Constraint from class-III observables $Br(B \to X_s \mu^+ \mu^-$) (left), $\tilde{A}_{FB}$ (middle) and $\tilde{F}_L$ (right)
at $1\, \sigma$ in the $(\delta C_7,\delta C_{7^\prime})$ plane in Scenario A together with the three (black) regions allowed by class-I observables. The magenta circle (left) and area (right) indicate the two regions where the three constraints are all satisfied.}
\label{fig:scenarioA}
\end{figure}

We start our analysis by considering the constraints on $(C_7,C_{7'})$ from the three class-I observables (left panel of fig.~\ref{fig:C7C7p}). They overlap on three (black) regions: one lies around the SM value, whereas two ``inverted'' ones are located where $C_7$ vanishes and $|C_{7'}|$ is of the same magnitude as $|C_7^{SM}|$. Despite the lesser theoretical control on the isospin asymmetry $A_I$ in $B\to K^*\gamma$, this observable proves interesting in discarding the so-called ``flipped-sign'' solution $(C_7,C_{7'})=(-C_7^{SM},0)$ discussed some time ago in connection with the apparent lack of zero in the $B\to K^*\ell^+\ell^-$ forward-backward asymmetry $A_{FB}$~\cite{Gambino:2004mv}. From our classification of observables, we know that these constraints on $(C_7,C_{7'})$ will hold for all the three scenarios discussed to be discussed now.

\begin{figure}
\includegraphics[width=4.99cm]{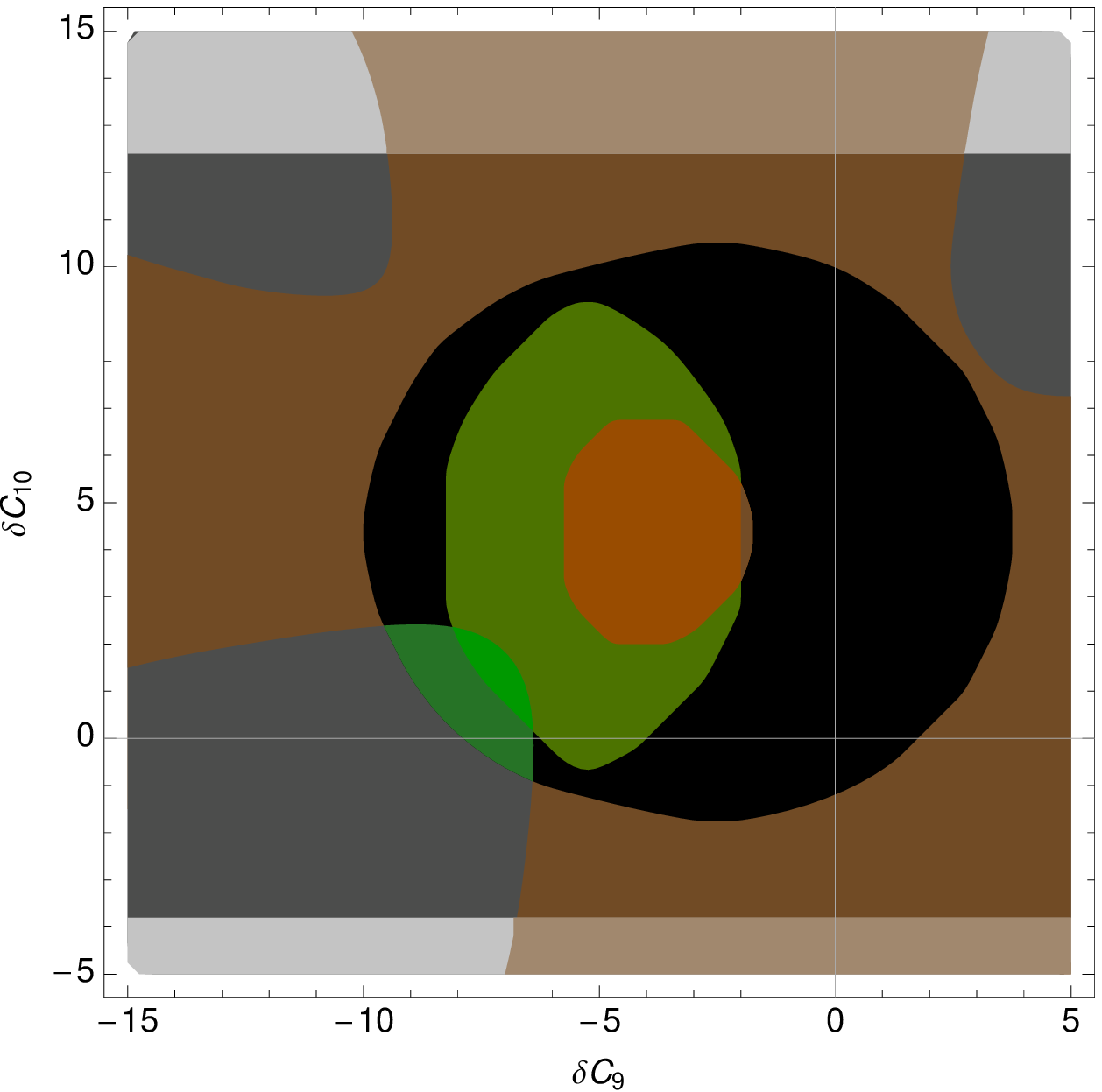}
\includegraphics[width=4.99cm]{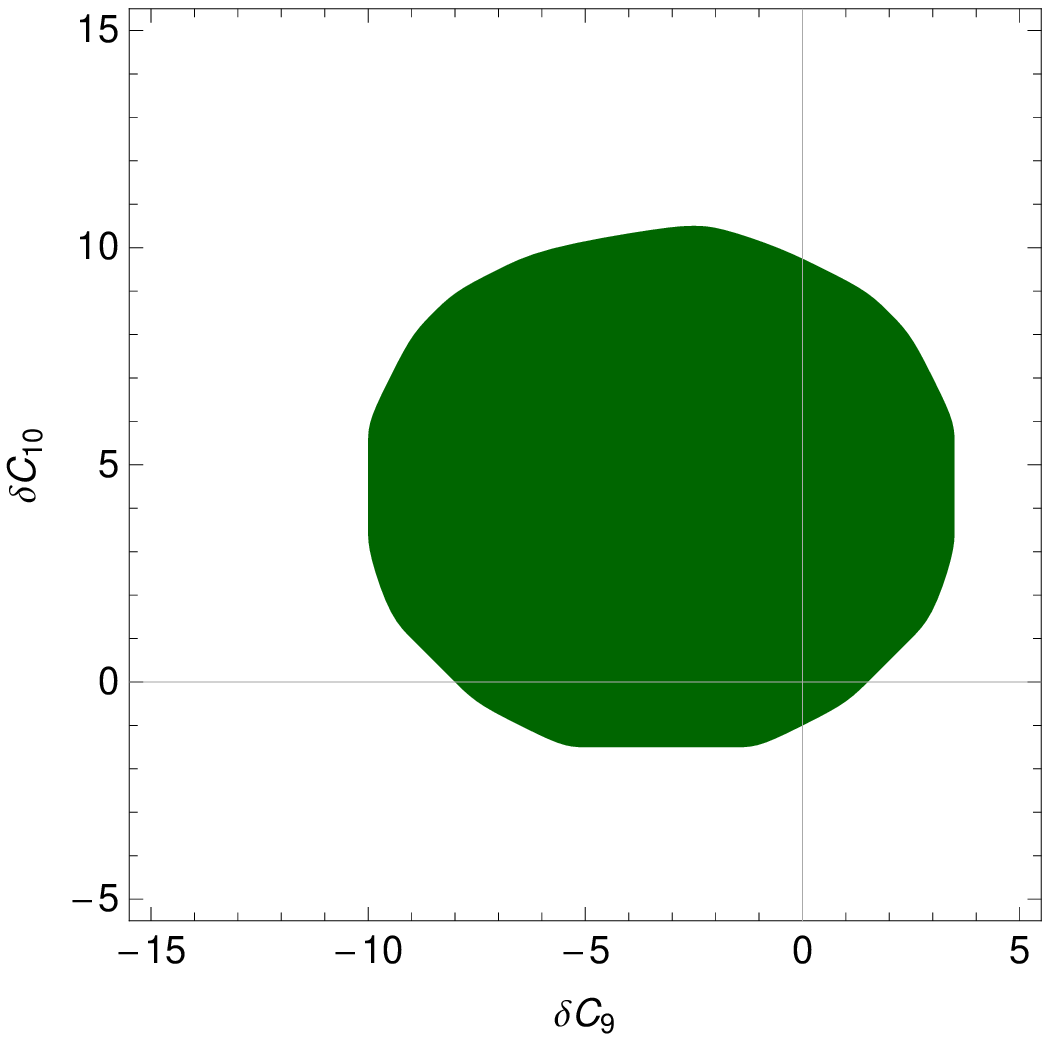}
\includegraphics[width=4.99cm]{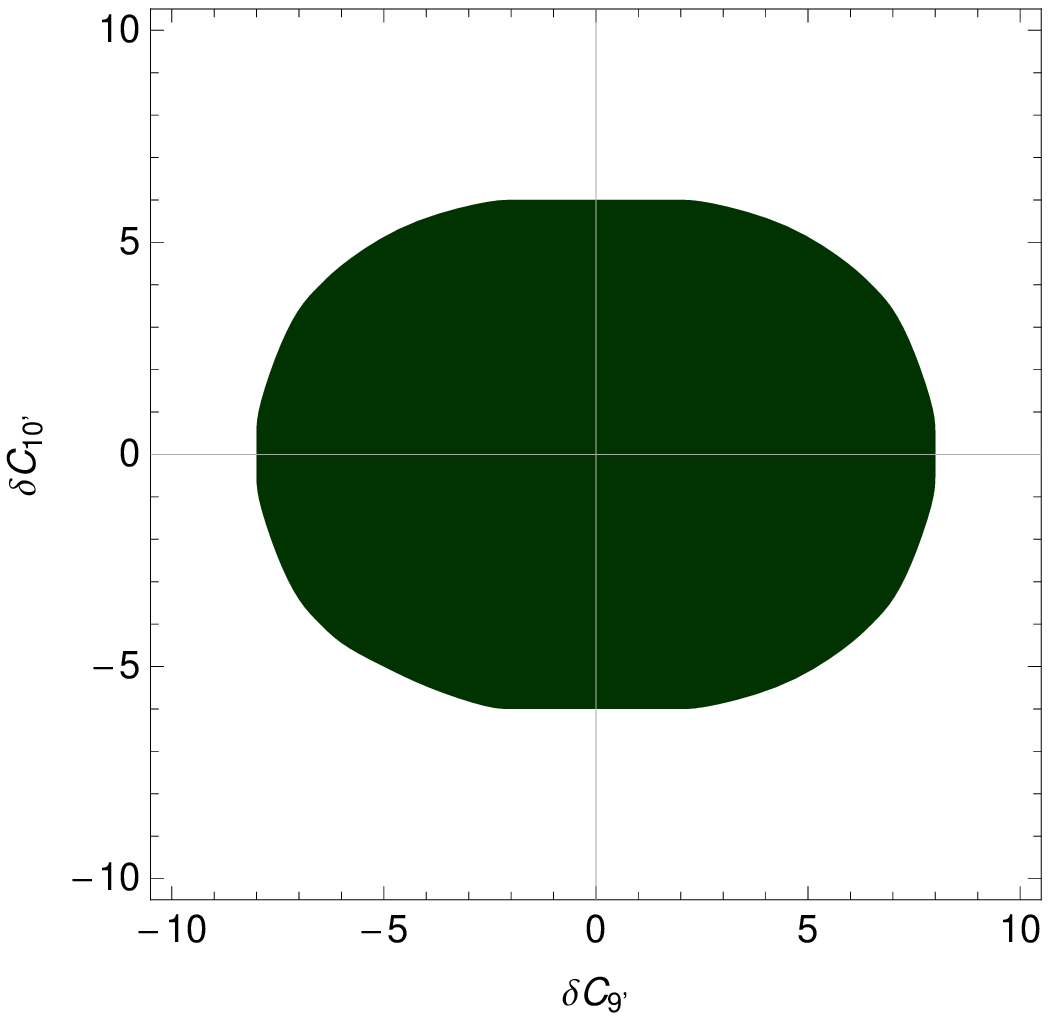}
\caption{On the left: overlap of the constraints from class-III observables $Br(B \to X_s \mu^+ \mu^-$) (green ring with an excluded central red region), $A_{\mathrm{FB}}$ (brown cross) and $F_{\mathrm{L}}$ (dark gray area with a central inlet) at $1\, \sigma$ in the $(\delta C_9,\delta C_{10})$ plane in Scenario B. The constraints imposed by their intersection are shown as a black crescent.
In the middle and on the right: constraints from class-III observable $Br(B \to X_s \mu^+ \mu^-$) at $1\, \sigma$ in the $(\delta C_9,\delta C_{10})$ and $(\delta C_{9^{\prime}},\delta C_{{10}^{\prime}})$ planes in Scenario C. In both scenarios, the regions shown are compatible with the constraints on $\delta C_7$ and $\delta C_{7^{\prime}}$ imposed by class-I observables on the left panel in fig.~1.}\label{fig:scenarioBC}
\end{figure}

\emph{In Scenario A} (NP only in electromagnetic dipole operators), 
the class-III observables constrain $C_7$ and $C_{7'}$ further, as shown in fig.~\ref{fig:scenarioA}. Their overlap selects a region around the SM-like solution [$(C_7,C_{7'})\simeq (C_7^{SM},0)$, magenta region on the right panel] and another at the edge of the ``inverted'' lower region [$(C_7,C_{7'})\simeq (0,C_{7}^{SM})$, circle on the left panel]. The two regions can be distinguished by the $q^2$-variation of the $B\to K^*\ell^+\ell^-$ transverse asymmetry $A_T^{(2)}$ (right panel of fig.~\ref{fig:C7C7p}) which almost vanishes at moderate $q^2$ for the SM-like solution and becomes negative for the ``inverted'' one.
\emph{In Scenario B} (NP affecting also semileptonic operators), the class-III observables turn out to constrain only $C_9,C_{10}$, as shown in the left panel of fig.~\ref{fig:scenarioBC}. $C_{10}$ is also bound by the updated upper limit on $Br(B_s\to\mu^+\mu^-)$, indicated by the light horizontal bands on the same plot. The overlap of all constraints yield a (black) region in the $(C_9,C_{10})$ plane, whereas $(C_7,C_{7'})$ must remain in the three (black) regions on the left-hand side of fig.~\ref{fig:C7C7p}. 
\emph{In Scenario C} (presence of right-handed semileptonic operators in addition to the previous contributions), only $Br(B\to X_s\mu^+\mu^-)$ turns out to be relevant in the $(C_9,C_{10},C_{9'},C_{10'})$ space due to its quadratic structure that translates into elliptic constraints in each plane, as indicated in the middle and right panels of fig.~\ref{fig:scenarioBC}. Unfortunately, in both scenarios B and C, the large uncertainty on semileptonic WCs leads to  values of $A_T^{(2)}$ spanning all its potential range, so that no firm prediction can be achieved.

An extension of the present analysis is planned, relying on a more careful statistical analysis, considering the $q^2$-variation of the $B\to K^*\ell^+\ell^-$ observables rather than their integrated values, and including a larger set of observables and NP operators (scalar, tensors\ldots).

\begin{thebibliography}{99}
 \bibitem{DescotesGenon:2011yn}
  S.~Descotes-Genon, D.~Ghosh, J.~Matias, M.~Ramon,
  JHEP {\bf 1106 } (2011)  099, and coming addendum.
  
\bibitem{QCDF}
  A.~L.~Kagan and M.~Neubert,
  Phys.\ Lett.\  B {\bf 539}, 227 (2002)
  [arXiv:hep-ph/0110078].
  
  T.~Feldmann and J.~Matias,
  JHEP {\bf 0301}, 074 (2003)
  [arXiv:hep-ph/0212158].

 M.~Beneke, T.~Feldmann and D.~Seidel,
  Eur.\ Phys.\ J.\  C {\bf 41} (2005) 173
  [arXiv:hep-ph/0412400].

\bibitem{Gambino:2004mv}
  P.~Gambino, U.~Haisch and M.~Misiak,
  Phys.\ Rev.\ Lett.\  {\bf 94}, 061803 (2005)
  [arXiv:hep-ph/0410155].
  


\bibitem{Kruger:2005ep}

  F.~Kruger, J.~Matias,
  Phys.\ Rev.\  {\bf D71 } (2005)  094009.
  [hep-ph/0502060].

 U.~Egede \emph{et al.},  
  JHEP {\bf 0811 } (2008)  032.
  [arXiv:0807.2589 [hep-ph]].

  
  \end{thebibliography}
\end{document}